\documentclass[11pt,titlepage]{article}

\usepackage{setspace} 

\usepackage{amsmath}

\usepackage{graphicx}

\usepackage[round,numbers,sort&compress]{natbib} 
\title{Molecular motors transporting cargos in viscoelastic cytosol: \\
how to beat subdiffusion with a power stroke?
}

\author{Igor Goychuk\thanks{
           Corresponding author.  Address: 
           Institute for Physics and Astronomy,
	   University of Potsdam,
	   Karl-Liebknecht-Str. 24/25,
	   14476 Potsdam-Golm, Germany, e-mail: igoychuk@uni-potsdam.de,
	   Tel.:~(049)331-977 5614, Fax:~(049)331-977-1045}, \\
	   Vasyl O. Kharchenko \\Institute of Applied Physics\\
	Natl. Acad. Sci. Ukraine, Sumy, Ukraine,\\
	\and Ralf Metzler \\
	Institute for Physics and Astronomy, \\
	University of Potsdam,Karl-Liebknecht-Str. 24/25,\\
	   14476 Potsdam-Golm, Germany \\ \&
	Department of Physics, Tampere University of Technology, \\
	Korkeakoulunkatu 3,
33101 Tampere, Finland   
	   }

\date{}

\begin{document}

\maketitle

\abstract{
Anomalously slow passive diffusion,  $\langle \delta x^2(t)\rangle\simeq t^{\alpha}$,
with $0<\alpha<1$, of larger tracers such as messenger RNA and
endogenous submicron granules in the cytoplasm of living biological cells has
been demonstrated in a number of experiments and has been attributed to the
viscoelastic physical nature of the cellular cytoplasm. This finding provokes
the question to which extent active intracellular transport is affected by this
viscoelastic environment: does the subdiffusion of free submicron cargo such as 
vesicles and organelles always imply anomalously slow transport by
molecular motors such as kinesins, that is, directed transport characterized
by a sublinear growth of the mean distance, $\langle x(t)\rangle\simeq t^{\alpha_{\rm eff}}$,
with $0<\alpha_{\rm eff}<1$? Here we study a generic model approach combining the commonly
accepted two-state Brownian ratchet model of kinesin motors based on the
continuous-state diffusion along microtubule driven by a flashing binding potential.
The motor is elastically coupled to a cargo particle, which in turn is subject to
the viscoelastic cytoplasmic environment. Depending on the physical parameters of
cargo size, loading force, amplitude of the binding potential, and the turnover
frequency of the molecular motor, the transport can be both normal ($\alpha_{\rm eff}=1$)
and anomalous ($\alpha\leq \alpha_{\rm eff}<1$). In particular, we demonstrate in detail how highly
efficient normal motor transport can emerge despite the anomalously slow passive
diffusion of cargo particles, and how the active motion of the same motor in the
same cell may turn anomalously slow when the parameters are changed.\\

}

\clearpage

\section*{Introduction}

Molecular transport inside biological cells comprises both passive and active processes \citep{Nelson}.
Thermal Brownian motion presents a ubiquitous mechanism especially for small particles like metal ions, 
aminoacids, sugar molecules, and even for larger particles like transcription
factors, enzymes, RNAs, lipid granules, etc. \citep{Luby}. Mean squared 
distance covered by a particle moving randomly without any mean bias, $\langle \mathbf{r}\rangle=0$, 
scales linearly with time,
$\langle (\delta \mathbf{r})^2\rangle=\langle \mathbf{r}^2\rangle-\langle \mathbf{r}\rangle^2 \propto D t$. 
If to apply a biasing external force $\mathbf{f}_{\rm ext}$, the 
mean distance in the direction of bias scales also linearly with time, 
$\langle \mathbf{r}\rangle \propto \mathbf{f}_{\rm ext} t/\eta$. The diffusion coefficient $D$, 
viscous friction coefficient
$\eta$, and temperature $T$ are related by the Einstein relation, $D=k_B T/\eta$ 
($k_B$ is the Boltzmann constant). It manifests
classical fluctuation-dissipation theorem (FDT) at local thermal equilibrium \cite{Kubo}. 
The Stokes formula, $\eta=6\pi \zeta a$, relates viscous friction with the medium's viscosity $\zeta$ and
radius $a$ for spherical particles in simple fluids. The larger the particle the smaller its
mobility $\mu=1/\eta$ and diffusion coefficient. In complex molecularly crowded polymeric fluids like cytosol 
this simple linear dependence
on the particle's size generally breaks down and the effective friction 
coefficient $\eta_{\rm eff}$ can exponentially be 
enhanced in cytosol with respect to one in water by many orders of magnitude, depending on the 
particle size, correlation length of polymeric fluid, 
and other parameters \citep{Odijk,Masaro}. 
If for a calcium ion the felt cytosol viscosity can be essentially the same as one of water, 
a vesicle or magnetosome with $a>100$ nm can feel it rather  as one of glycerol 
($1500\times$ more viscous than
water at room temperatures), or even honey ($10000\times$ more viscous), as it can be guessed
from recent systematic studies in  \citep{Holyst} for a more simple system. 
Even for particles of a typical size of globular protein $a\sim 2.5$ nm the enhancement factor 
can be as large as $750$, as derived in \citep{GoychukPRE12} from the experimental
data in \citep{Guigas}. 


Passive diffusion of submicron particles in living cells slows
down tremendously, if to consider it as a normal diffusion process on macroscale,
with respect to one in pure water. However, even this is not necessarily the
major effect because cytosol behaves as a viscoelastic polymeric fluid displaying profound
memory effects. Viscoelasticity alone can cause a subdiffusive behavior,
$\langle (\delta \mathbf{r})^2\rangle \propto D_{\alpha} t^\alpha$, with $0<\alpha<1$
and subdiffusion coefficient $D_{\alpha}$,
which has already been found  in
numerous experiments
\citep{Guigas,Amblard,Saxton,Qian,Yamada,Caspi,Tolic,Golding,Szymanski,Robert,Bruno,Jeon11,Jeon,
Barkai,Tabei,Hofling,Taylor,WeissM}, though other subdiffusive 
mechanisms were also suggested to explain the experimental results \citep{Barkai,Hofling}. 
They can also emerge in a combination with viscoelasticity, as the results in \citep{Jeon11,Tabei} suggest. 
In particular, crowding in polymer liquids has been shown to cause namely 
viscoelastic subdiffusion \citep{Szymanski,WeissM}, which is related to the 
fractional Brownian motion \cite{Mandelbrot}, as discussed in \citep{Goychuk09}. This connection
can be derived \cite{Wang,GH07} from a Generalized Langevin Equation, or GLE \citep{Kubo,Zwanzig}
with a power-law decaying memory kernel and fractional Gaussian thermal
noise \citep{Goychuk09,Goychuk12} as a dynamically well-founded approach with deep roots in
statistical mechanics \citep{Zwanzig,ZwanzigBook,KuboBook}. Cytosol is a highly crowded 
and viscoelastic liquid, which
is a common point \citep{Ellis,Mcguffee}. This approach is used also in this work.
As an example pertinent to this work,  the subdiffusive motion of magnesomes with $a=300$ nm 
and their chains (up to 8 magnetosomes in a chain) 
in Ref. \citep{Robert} is subdiffusive with $\alpha\approx 0.4$ in intact cytosol. It is
characterized by subdiffusion
coefficients as small as $D_{0.4}\sim 10^{-16}\;{\rm m^2/s^{0.4}}=100\;{\rm nm^2/s^{0.4}}$, 
and even smaller,
depending on the number of magnetosomes in the chain. Therefore, the corresponding 
subdiffusional spread within one second is of the order of 10 nm only. 
This brings the theme of active transport by molecular motors into the focus of attention.

Without help of molecular motors
such particles would be practically localized on appreciable long time scales. 
The motors are thus crucial for the delivery of such and similar cargos in living cells \citep{Hirokawa,Jones}.
The theory of molecular motors viewed as Brownian stochastic engines is well developed for 
memoryless Markovian dynamics only, in the complect neglection of non-Markovian memory effects
caused, in particular, by the viscoelasticity of cytosol. This necessitates that
transport by molecular motors
\citep{Chauwin,Julicher1,Julicher2,Astumian,AstumianBier,Fisher,Perez} in viscoelastic media
should be elaborated in basic detail. We started to do this in Refs.
\citep{Goychuk10,Goychuk12,GKh12a,KhG12,GKh13a,KhG13a}.
 The field of Brownian ratchets  \citep{Reimann} is allied  to molecular motors
\citep{Chauwin,Julicher1,Julicher2,Astumian,AstumianBier,Makhno}, though it is dealing first and
foremost with more general problems of statistical physics. Within a generalist model, one can
think of the transport as one realized by  the motor particle with a tightly coupled cargo
making one compound particle moving in a periodic  external force field provided by its
interaction with microtubule (in the case of kinesins), which depends on the conformation
of motor protein particle. 
A corresponding generalization of the standard continuous diffusion ratchet model of molecular motors
towards viscoelastic subdiffusion has been put forward recently in \citep{GKhMetz13a}.
It explains a number of experimental facts, in particular, that the transport by molecular 
motors in viscoelastic cytosol can be both normal and anomalously slow, depending in particular
on the motor operating frequency and the cargo size.

In the present work, we provide a further  generalization of this recent model  in
\citep{GKhMetz13a} by considering transport of large subdiffusive cargos attached on elastic
linkers to the motors. Here, the assumption of absolutely rigid linker between the motor
and its cargo is relaxed.  Similar
models have been considered earlier for the normal diffusion of both cargo and motor
\citep{Zeldovich}. The dynamics of motor is also normal  in this paper, without
memory effects, when it is left alone. We characterize it by a largely 
reduced (by a factor of 10) diffusion
coefficient with respect to one expected in water. However, the coupling to 
subdiffusive cargo enforces the motor's
subdiffusion, when it is not coupled to a microtubule.  Considering  experimentally
relevant  elastic constants of  linker \citep{Bruno,Kojima} and other realistic parameters we
confirm all the major features revealed in \citep{GKhMetz13a} in a more general 
setup. The anomalous transport regime
becomes, however, reinforced. It emerges already for
motor turnover frequencies of the order of 100 Hz for sufficiently large cargos of a typical size
$a=300$ nm. However, even for such large cargos the transport can become normal if turnover
frequency is lowered to 10 Hz. Generally, the dependence of the effective subdiffusive transport
exponent  on the cargo size and motor turnover frequency would make a decisive
test in favor of our theory of anomalous transport mediated by molecular motors in living cells.

\section*{Model and theory}

Diffusion in such complex viscoelastic fluids as cytosol is commonly described by 
the Generalized Langevin Equation, or GLE \citep{Mason,Waigh}. We
consider it here for overdamped dynamics, in neglection of inertial effects, 
written for one Cartesian coordinate $y$ for simplicity,
\begin{eqnarray}
\label{GLE}
\int_{-\infty}^t\eta_c(t-t')\dot y(t')dt' =f_{\rm ext}(t) +\xi_c (t) \;.
\end{eqnarray}
 Here the memory kernel $\eta_c(t)$ and the 
autocorrelation function of unbiased thermal 
colored Gaussian noise $\xi_c(t)$ are related by the second FDT of Kubo, named also 
fluctuation-dissipation relation (FDR),
\begin{eqnarray}
\label{FDR}
 \langle \xi_c(t)\xi_c(t')\rangle=k_BT\eta_c(|t-t'|)\;.
\end{eqnarray}
It reflects the energy balance at thermal equilibrium between the energy pumped by thermal noise and 
energy dissipated due to friction. Such a stochastic description is not only consistent with the laws of 
equilibrium statistical physics and thermodynamics, but it 
also allows to treat strongly out-of-equilibrium
transport, driven e.g. by a non-thermal fluctuating force $f_{\rm ext}(t)$.  GLE~\ref{GLE}
serves as a basis in passive microrheology \citep{Mason,Waigh,Qian}, at thermal equilibrium,  
to derive the complex shear 
modulus of the medium in the frequency domain, 
$G^*(\omega)\propto i\omega\int_0^{\infty}\exp(-i\omega t)\eta(t)dt$,  from the particle trajectories. 
The complex shear modulus is commonly used to characterize viscoelastic materials \citep{JonesBook}.
In particular, a frequently observed 
power law scaling $G^*(\omega)\propto (i\omega)^\alpha$ with $0<\alpha<1$ corresponds
to subdiffusion,  $\langle (\delta x)^2\rangle \propto D_{\alpha} t^{\alpha}$,  with fractional
diffusion coefficient $D_\alpha$ and power law
scaling of memory decay, $\eta(t)\propto \eta_\alpha/t^\alpha$, where $\eta_\alpha$ is fractional 
friction coefficient obeying fractional Einstein relation $D_\alpha=k_BT/\eta_\alpha$ \cite{Goychuk12}.
In macroscopic theory of viscoelasticity, 
similar memory kernels were introduced long ago by A. Gemant \citep{Gemant} as a generalization
of the simplest Maxwell model with exponentially decaying memory \cite{Maxwell}.
 In practice,
such a power law scaling extends mostly over several time and frequency decades. 
High-frequency (short-memory) 
cutoff reflects molecular nature of the condensed medium. A low-frequency, or 
long-memory cutoff guarantees that
the macroscopic friction coefficient $\eta_{\rm eff}=\int_0^\infty \eta(t)dt$ is finite, which reflects 
finite viscosity of any fluid on macroscale \cite{Goychuk09,Goychuk12}. Intermediate power law scaling
gives rise to subdiffusion on a transient time scale (up to several minutes, depending
on the particle's size), 
and this can establish subdiffusion as a primary
passive transport mechanism for submicron particles on the mesoscale of biological cells interior.

The mathematical model of a strictly algebraically decaying  memory kernel corresponds 
to the fractional Gaussian
noise (fGn) model of thermal noise.  fGn presents a time-derivative of the fractional Brownian motion
(fBm) \citep{Mandelbrot}.  Both can be characterized by the Hurst exponent $H'=1-\alpha/2$. 
Such a noise is persistent for $1/2<H'<1$, with positive correlations. It corresponds
to the sub-Ohmic model of thermal baths consisting of harmonic oscillators \cite{Weiss}. 
The corresponding GLE can be derived from 
a purely dynamic hyper-dimensional Hamiltonian model assuming merely initial canonical distribution 
of thermal bath
oscillators at a given temperature, like in a typical molecular dynamics setup \cite{Zwanzig,Weiss}. 
It has thus the firm
statistic-mechanical foundation.
The solution of GLE \ref{GLE} is then also fBm, but anti-persistent and subdiffusive, with the Hurst exponent
$H=\alpha/2$. This transformation occurs due to the friction with algebraically 
decaying memory \citep{GH07}. Important, this is namely the memory friction
which is at the heart of the very phenomenon of viscoelasticity.
Experimental values of $\alpha$ can be very different, in the range of
$\alpha = 0.2\div 1$ \citep{Hofling}. For example, for the intact cytoskeleton
in Ref. \citep{Robert}, $\alpha=0.4$. About the same value can be derived from the experimental data in 
Ref. \citep{Bruno}, namely from the power spectrum $S(\omega)$ of the transversal position fluctuations
of the melanosome particles (size $a=250$ nm) elastically attached to the motor proteins 
walking along microtubule.
For sufficiently large frequencies (exceeding inverse relaxation time scale
in parabolic potential well), $S(\omega)\propto 1/\omega^{b}$, with $b=1+\alpha$ and 
experiment yields $b=1.41\pm 0.02$. We accept $\alpha=0.4$ as 
an experimentally relevant numerical value in this work. 

\subsection*{Earlier modeling}

The simplest idea to model the influence of molecular motors on the  cargo dynamics within the
GLE approach is to approximate their  collective influence  by a time-dependent random force
$f_{\rm ext}(t)$,  which itself exhibits a long range memory and is power law correlated,
$\langle f_{\rm ext}(t') f_{\rm ext}(t)\rangle \propto 1/|t-t'|^\gamma$, with  $0<\gamma<1$.
This can lead to  superdiffusion \citep{BrunoPRE},  $\langle (\delta
y^2\rangle\propto t^\beta$, with $\beta=2\alpha-\gamma>1$  \citep{Wang,BrunoPRE},
for $\alpha>0.5$. Notice, however, that $\beta$ can take the maximal 
value of $2\alpha$, for $\gamma\to 0$, within this model. This corresponds to a strict $1/f$ noise
driving force  $f_{\rm ext}(t)$ generated by motors with almost non-decaying correlations. 
The origin of this  limit can
easily be understood, if to consider the transport by a time-alternating force $\pm f_0$ in
the opposite directions.  Then, $\langle \delta y(t)\rangle\propto \pm f_0 t^\alpha$, for a
force realization,  and after averaging over the driving force fluctuations $\langle\langle
\delta y(t)\rangle^2\rangle_{f_0}\propto t^{2\alpha}$. Any decaying correlations of the alternating
driving force make the effective diffusion exponent  smaller than $\beta_{\rm max }=2\alpha$.
Clearly, within such a description superdiffusion caused by  the activity of molecular motors
could only be possible for $\alpha>0.5$, at odds with experimental results showing $\beta\sim
1.2\div 1.4$ for $\alpha=0.4$ \citep{Robert}. Such  a model
is therefore too simple. It cannot explain some important experimental findings
consistently, providing but a useful first insight. 

\subsection*{Our model}

We consider thus a further generalization, where the cargo is elastically coupled to the motor with
coordinate $x$ (one-dimensional model), and $f_{\rm ext}=k_L(x-y)$, with spring constant $k_L$.
The motor undergoes diffusion, generally characterized by a memory friction $\eta(t)$, 
on microtubule in a binding potential $U(x,\zeta(t))$ reflecting spatial 
period $L$ of microtubule. The potential depends on the dynamical motor conformation $\zeta(t)$,
the motor's internal degree of freedom. Our model reads,
\begin{eqnarray}
\label{model1}
\int_{-\infty}^t\eta_c(t-t')\dot y(t')dt' & = &-k_L(y-x) +\xi_c (t),\\
\int_{-\infty}^t\eta(t-t')\dot x(t')dt' & = &
k_L(y-x)-\frac{\partial}{\partial x}U(x,\zeta(t))\nonumber \\ && -f_0 +\xi(t), \label{model1_2}
\end{eqnarray}
where the memory kernels  and the autocorrelation functions of unbiased thermal 
colored Gaussian noises are related by FDRs \ref{FDR}, 
and $\langle \xi(t)\xi(t')\rangle  = k_BT\eta(|t-t'|)$; $\xi_c (t)$, and $\xi(t)$
are not correlated. $f_0$ in \ref{model1_2} is a constant loading external force which can 
oppose the fluctuation-induced  transport and stop it.
We assume further that  dynamical conformation $\zeta(t)$
can take two values $\zeta_1$ and $\zeta_2$. The binding potential is periodic,
 $U(x+L,\zeta_{1,2})=U(x,\zeta_{1,2})$, but spatially asymmetric, see in Fig. \ref{Fig1}, lower inset.
 Spatial periodicity guarantees that transport is absent in the absence of conformational
 fluctuations induced by ATP binding to molecular motor and its hydrolysis. It is forbidden
 by the symmetry of thermal detailed balance, at thermal equilibrium, in the absence
 of an energy source.  We assume that two consequent conformational switches
 make one cycle, and 
$U(x+L/2,\zeta_{1})=U(x,\zeta_{2})$. Switching between two conformations is considered as symmetric
two-state Markovian process with the equal transition rates $2\nu_{\rm turn}$.

To be more specific, one considers piecewise linear sawtooth potential (piecewise constant potential force)
with amplitude $U_0$ and period $L$. The minimum divides the period in the ratio $p:1$, and
we take a particular value $p=3$. This asymmetry defines
the natural direction of transport towards positive $x$.  The maximally possible loading, or 
stalling force for this potential at zero temperature is easy to
deduce:
$f_0^{stop}(T=0)=(p+1)U_0/(pL)$. In units of 
room $k_BT_r=4.1\cdot 10^{-21} J=4.1 \;{\rm pN\cdot nm}$,  $L=8$ nm, 
and $f_0^{stop}(T=0)\approx 0.6833\; U_0/(k_BT_r)$ pN  for $p=3$. 
For $T=T_r$ it will be essentially lower, see below. 
We choose $U=20\;k_BT_r$ in our simulations. This corresponds to $f_0^{stop}(T=0)=13.67$ pN, about twice
the maximal loading force of kinesin II at physiological temperatures.

In this work, we neglect 
the memory effects for the motor particle, $\eta(t)=2\eta_m\delta(t)$, with Stokes friction
$\eta_m=6\pi a_m\zeta_w$, where $a_m$ is the effective radius of motor molecule and 
$\zeta_w=1\;{\rm mPa\cdot s}$. We take $a_m=100$ nm, about 10 times larger than  a linear geometrical size
of kinesin in order to account for the enhanced effective viscosity felt by the 
motor in cytosol with respect to its water constituent.
A characteristic time scale $\tau_m=L^2\eta_m/U_0^*$ has been used for scaling time in numerical simulations,
with $U_0^*=10\;k_BT_r$.
For the stated parameters, $\tau_m\approx 2.93\;\mu{\rm s}$. Distance is scaled in units of $L$, and 
elastic coupling constants in units of $U_0^*/L^2\approx 0.64$ pN/nm. Next, we single out purely
viscous component in the memory kernel $\eta_c(t)=2\eta_c\delta(t)+\eta_{\rm mem}(t)$, where
$\eta_{\rm mem}(t)=\eta_\alpha/(\Gamma(1-\alpha)t^\alpha)$. 

If the cargo is uncoupled from the motor ($k_L\to 0$), its diffusional behavior is described by 
\citep{KhG13a}
\begin{eqnarray}\label{exact}
\langle \delta y^2(t)\rangle = 2D_c tE_{1-\alpha,2}\left (-(t/\tau_{\rm in})^{1-\alpha}\right )\;,
\end{eqnarray}
where $E_{a,b}(z)=\sum_{n=0}^\infty z^n/\Gamma(an+b)$ is generalized Mittag-Leffler function, 
$D_c=k_BT/\eta_c$ is normal diffusion coefficient, and 
$\tau_{\rm in}=(\eta_c/\eta_\alpha)^{1/(1-\alpha)}$ is a time constant separating initially
normal diffusion, $\langle \delta y^2(t)\rangle\approx 2D_c t$ at $t\ll\tau_{\rm in}$, and
subdiffusion, $\langle \delta y^2(t)\rangle\approx 2D_\alpha t^\alpha/\Gamma(1+\alpha)$ for 
$t\gg\tau_{\rm in}$. Furthermore, in accordance with the methodology in \citep{Goychuk09,Goychuk12} 
we approximate the memory kernel by a sum of exponentials 
\begin{eqnarray}
\label{memory}
\eta_{\rm mem}(t)\approx \eta_{\rm mem}(t,\nu_0,b,N)=\sum_{i=1}^{N}k_i\exp (-\nu_i t),
\end{eqnarray}
obeying a fractal scaling, $\nu_i=\nu_0/b^{i-1}$, $k_i\propto \nu_i^\alpha$.
 This form ensures a power law scaling, $\eta_{\rm mem}(ht,\nu_0,b,N)=h^{-\alpha}\eta_{\rm mem}(t,h\nu_0,b,N)$,
for $t$ being well within the time-interval between two cutoffs, $\tau_{\rm min}=\nu_0^{-1}$ and 
$\tau_{\rm max}=\tau_{\rm min} b^{N-1}$. The dilation parameter $b>1$ controls the accuracy of
approximation, which exhibits small logarithmic oscillations with respect to the exact
power law, and $\nu_0$ corresponds to a high-frequency cutoff reflecting atomic/molecular
nature of any physical condensed environment. 
In this respect, any physical fractal, either spatial or in time, has minimal 
and maximal ranges, defined in our case by the ratio $\tau_{\rm max}/\tau_{\rm min}=b^{N-1}$, 
and even a rough decade scaling with $b=10$ allows  to approximate 
the power law with accuracy of several percents for $\alpha=0.4$ (further
increased to a one hundredth of percent for $b=2$ \citep{GKh13a}). 
Similar expansions are well known in the theory of anomalous relaxation \cite{Palmer}.
This provides a foundation for
excellent numerical method to integrate fractional GLE dynamics. 
Alternatively, this methodology can be considered as an independent approach to model anomalous
diffusion and transport on physically relevant spatial and time scales, not caring much about
its relation to fractional GLE. 
Then, a  convenient parameterization is 
\begin{eqnarray}
k_i=\nu_0\eta_{\rm eff}\frac{b^{1-\alpha}-1}{b^{(i-1)\alpha}[b^{N
(1-\alpha)}-1]}\;,
\end{eqnarray}
where $\int_0^{\infty }\eta_{\rm mem}(t)dt=\eta_{\rm eff}$ characterizes largely enhanced
macroscopic friction coefficient in a long-time limit, $t\gg \tau_{\rm
max}$. The number $N$ 
of auxiliary quasi-particles controls the maximal
range of subdiffusive dynamics, which becomes again normal,
$\langle \delta y^2(t)\rangle \sim 2 D_{\rm eff} t$, for  $t\gg
\tau_{\rm max}$ with $D_{\rm eff}=k_BT/(\eta_{\rm
eff}+\eta_c)$.   Notice that 
the fractional friction coefficient is related to the effective friction 
as $\eta_{\alpha}=\eta_{\rm eff}\tau_{\rm
max}^{\alpha-1}/r$. Here,  $r=\frac{C_\alpha(b)}{\Gamma(1-\alpha)}
\frac{b^{1-\alpha}}{b^{1-\alpha}-1}[1-b^{-N(1-\alpha)}]$ is a
numerical coefficient of the order of unity,  $r\approx 0.93$ for $N\geq 5$,
at $\alpha=0.4$, and $b=10$, with $C_{0.4}(b)\approx 1.04$.
 The effective relative friction coefficient
$\tilde \eta_{\rm eff}= \eta_{\rm eff}/\eta_c$ is used as an important parameter in
our simulations. It defines the time range of subdiffusion, from $\tau_{\rm
in}=\tau_{\rm
max}/\tilde\eta_{\rm eff}^{1/(1-\alpha)}$ to $\tau_{\rm max}$. For example, for $\tilde\eta_{\rm eff}=10^3$
and $\alpha=0.4$ one expects that subdiffusion will extend over 5 time decades.
According to this important, nontrivial but simple relation, the relative increase of effective
friction controls the relative time range of subdiffusion 
$\tilde\eta_{\rm eff}^{1/(1-\alpha)}$, independently of $b$, and $N$!

The discussed approximation allows to replace
the non-Markovian GLE dynamics with a higher dimensional Markovian dynamics
upon introduction of the  $N$ auxiliary Brownian quasi-particles with coordinates $y_i$ accounting for
viscoelastic degrees of freedom, see in \cite{Goychuk12} for details. In the
present case,
\begin{eqnarray}
\label{embedding}
\eta_m \dot x & =& f(x,\zeta(t))-k_L(x-y)+\sqrt{2\eta_m k_BT}\xi_m(t) \;,\nonumber \\
\eta_c \dot y & =& k_L(x-y)-\sum_{i=1}^{N}k_i(y-y_i)+\sqrt{2\eta_c k_BT}\xi_0(t) \;,\nonumber \\
\eta_i\dot y_i& = &k_i (y-y_i)+\sqrt{2\eta_ik_BT}\xi_i(t) \;,
\end{eqnarray}
where $f(x,\zeta(t))=-\partial U(x,\zeta(t))/\partial x-f_0$, and $\eta_i=k_i/\nu_i$. 
Furthermore, $\xi_i(t)$ are uncorrelated
white Gaussian noises of unit intensity, $\langle \xi_i(t')\xi_j(t)\rangle=
\delta_{ij}\delta(t-t')$, which are also uncorrelated with white Gaussian noises
 $\xi_0(t)$ and $\xi_m(t)$. To have a
complete equivalence with the stated GLE model in Eqs. \ref{model1}-\ref{memory}, the initial positions
$y_i(0)$ are sampled from
a Gaussian distribution centered around $y(0)$, $\langle y_i(0)\rangle=y(0)$
with variances $\langle[y_i(0)-y(0)]^2\rangle=k_BT/k_i$.
The stochastic variable $\zeta(t)$ is described by a symmetric two-state Markovian process with the
identical transition rates $\nu=2\nu_{\rm turn}$. This corresponds to the simplest
model of molecular motors considered as flashing ratchets \citep{Nelson,Chauwin,AstumianBier}. 
By doing numerical solutions of the
set \ref{embedding} with a time step $\delta t\ll \nu^{-1}$, the variable 
$\zeta(t)$ alternates its state with the
probability $\nu\delta t$ at each integration time step, or continues to stay in the same
state with probability $1-\nu\delta t$. This  is decided upon comparison  of  a
(pseudo)-random number uniformly distributed between zero and one with $\nu\delta t$.

\subsection*{Thermodynamic efficiency and energetic efficiency of the cargo delivery}

Thermodynamic efficiency of anomalous Brownian motors has been addressed quite recently \citep{GKh13a,KhG13a}.
It turns out that the general approach developed earlier for normal motors 
\citep{Sekimoto,Julicher1} can be almost straightforwardly applied. It  yields, however, several important
 new results in the anomalous transport regime. Generically, the work done by the potential fluctuations
induced by the enzyme turnovers, or the input energy pumped into directed 
motion is
$E_{\rm in}(t)=\int_0^t\frac{\partial }{\partial t} U(x,t) dt$ \citep{Sekimoto}. 
It can be further calculated, within the considered model, 
as a sum of potential energy jumps $\Delta U(x(t_i))$ occurring at random instants of time
$t_i$ marking cyclic conformational 
transitions $1\to 2\to 1\to ...$. 
This input energy is spent on doing work against external loading force $f_0$.
The corresponding useful work is $W_{\rm use}(t)=f_0\delta x(t)$.
The rest is used to overcome the dissipative influence of environment. It is dissipated as heat. 
The thermodynamic efficiency of isothermal motors is
thus just $R_{\rm th}(t)=\langle W_{\rm use}(t)\rangle/\langle E_{\rm in}(t)\rangle$, 
upon averaging over many ensemble realizations. Clearly, $R_{\rm th}=0$  for $f_0=0$. 
This is also very clear from an energetic point of 
view as by relocation from one place to another neither potential energy of cargo, nor that of motor has been
changed. This is a normal \textit{modus operandi} of such motors as kinesin which is very different from other
molecular machines such as ion pumps  which are primarily transferring ions against an electrochemical 
potential gradient, i.e. increase electrochemical potential of ions.
Anomalously slow transport introduces principally new features for $f_0>0$.
Namely, for anomalous transport the useful work done against $f_0$ scales sublinearly with
time, $\langle W_{\rm use}(t)\rangle\propto f_0 t^{\alpha_{\rm eff}}$, while the input energy
scales linearly, $\langle E_{\rm in}(t)\rangle\propto t$. It is proportional to the
mean number of potential fluctuations. By the some token as fractional transport
cannot be characterized by a constant mean velocity, it cannot be also characterized by mean power of useful work
in a stationary
regime. However, one can define fractional power and fractional efficiency \citep{GKh13a,KhG13a}. 
Thermodynamic efficiency simply does not present a completely adequate measure in such a situation. 
Nevertheless, it decays algebraically slow in time, $R_{\rm th}(t)\propto 1/t^{1-\alpha_{\rm eff}}$
and can be still rather high even for large times, see below. Moreover, the dependence of $R_{\rm th}(t,f_0)$ 
on load $f_0$ is also very illustrative, $R_{\rm th}(t,f_0)$ vanishes not only at $f_0=0$, but also at 
some $f_0^{stop}(T,U_0,\nu_{\rm turn})$. It defines the stalling force, which is 
time-independent, but strongly
depends on the potential amplitude, temperature, and driving frequency, as will be shown
also below. Thermodynamic efficiency has thus a maximum at some optimal loading force 
$f_0^{opt}(t)$, and $R_{\rm th}(t,f_0^{opt})$ can be in anomalous transport regime still very high,
for a sufficiently high $U_0$, comparable with the maximal thermodynamic efficiency 
of kinesins in normal regime of about 50\%. The dependence of 
$R_{\rm th}(t,f_0)$ on $f_0$ is strongly asymmetric in a thermodynamically highly efficient regime. 
However, it becomes more symmetric in a low efficient anomalous regime, where it is described
approximately by a parabolic dependence, 
$R_{\rm th}\propto f_0(1-f_0/f_0^{stop})/f_0^{stop}$ with $f_0^{opt}=f_0^{stop}/2$, and a proportionality
coefficient which slowly drops in time.
This phenomenon of a time-dependence of $R_{\rm th}(t)$ 
can be verbalized as  fatigue of molecular motors caused by viscoelasticity
of cytosol. It can be considered as one of indications in the favor
of our theory if revealed experimentally. In the anomalous transport regime asymptotically
$R_{\rm th}(t)\to 0$ independently of $f_0$. Even though the useful work done against $f_0$ 
is always finite, which is a benchmark of any genuine Brownian motor, it becomes a negligible portion
of input energy in the course of time. The input energy is spent mostly to overcome the dissipative
influence of the environment with hugely enhanced effective viscosity, which is quite natural. 
It is dissipated as heat, $Q(t)=\langle E_{\rm in}(t)\rangle-\langle W_{\rm use}(t)\rangle$.

However, the primary utility of such motors as kinesin consists in delivery various cargos 
to certain destinations and
not in increasing their potential energy. For
this reason, numerous Stokes efficiencies have been defined in 
addition to $R_{\rm th}$ \citep{Derenyi,Suzuki,WangOster}. Which of them
is most appropriate remains dim, especially for anomalous transport \citep{KhG13a}.
We proposed a different
measure to quantify the motor performance at $R_{\rm th}=0$ named energetic delivery 
performance \citep{GKhMetz13a}.
It reflects optimization of the mean delivery velocity per energy spent. If to quantify the net input
energy in the number of enzyme turnovers (number of ATP molecules consumed as fuel), a natural
definition is  $D=d/(t \langle N_{\rm turn}\rangle)$, where $d$ is the delivery distance in time $t$, after $\langle N_{\rm turn}\rangle$ cyclic 
turnovers on average.
Clearly, $\langle N_{\rm turn}\rangle =\nu_{\rm turn} t$, and for an ideal motor in the tight coupling
power stroke regime, whose processive motion is perfectly synchronized with the turnovers of  ``catalytic wheel'' \citep{Wyman}, $d=L\nu_{\rm turn} t$, and therefore $D_{\rm ideal}=L^2\nu_{\rm turn}/d$,
i.e. for any given $d$, the increase of $\nu_{\rm turn}$ leads to a linearly increased delivery performance.
Clearly, in reality there will be always deviations from this idealization. With increasing turnover
frequency, or upon increasing load even normally operating motors have no enough time to relax down 
potential minimum after
each potential jump, or escape events become  important. This results in backsteps, and after reaching
a maximum versus $\nu_{\rm turn}$ the delivery efficiency will necessarily drop. In the anomalous
transport regime, the larger the delivery distance $d$ the smaller  the corresponding optimal 
$\nu_{\rm turn}^{opt}(d)$. Energetically, it makes than less sense to hydrolyze more ATP molecules 
for the efficient cargos delivery.  A corresponding optimization can be important in the cell economy.

\section*{Results}

We  use in our simulations $N=10$, $b=10$, and $\nu_0=100$ yielding
$\tau_{\rm max}\approx 29.4$ s for $\tau_m=2.94\;\mu{\rm s}$. 
Numerical solution of stochastic differential
equations \ref{embedding} was done implementing stochastic Heun method \citep{Gard} in CUDA
on NVIDIA Kepler graphical processor units. Time-step of integration was $\delta t=5\cdot 10^{-3}$
in the scaled units and the terminal time  was $10^6$ ($2.94$ s) in most simulations. We considered
a cargo with $a=300$ nm ($\eta_c/\eta_m=3$) and two different values of 
$\tilde \eta_{eff}=3\cdot 10^4$ (which we shall name ``larger'') and 
$\tilde \eta_{eff}=3\cdot 10^3$ (``smaller''). This corresponds
to two different values of subdiffusion coefficient 
$D_{\alpha}^{(1)}\approx 171\;{\rm nm^2/s^{0.4}}$, in 
accordance with typical values measured for magnetosomes 
in \citep{Robert} and ten times larger $D_{\alpha}^{(2)}\approx 1710\;
{\rm nm^2/s^{0.4}}$ which can be attributed to a smaller particle.
Furthermore, two different values of elastic spring constant were used $k_L^{(1)}=0.32$ pN/nm (``strong''),
which corresponds to measurements in vitro \citep{Kojima}, and a ten times softer 
$k_L^{(2)}=0.032$ pN/nm (``weak'') in accordance with the recent results in \citep{Bruno}, 
in living cells. Furthermore, $\nu_{\rm turn}^{(1)}=85$ Hz and $\nu_{\rm turn}^{(2)}=17$ Hz
denote two turnover frequencies (``fast'' and ``slow''). Five different sets of parameters, labeled
as $S_1,S_2,S_3,S_4,S_5$ correspond to ``larger cargo, stronger linker, fast'',
``smaller cargo, stronger linker, fast'', ``smaller cargo, weaker linker, fast'', 
``larger cargo, weaker linker, fast'', and ``larger cargo, stronger linker, slow'', respectively.

\begin{figure}
   \begin{center}
      \includegraphics*[width=3.5 in]{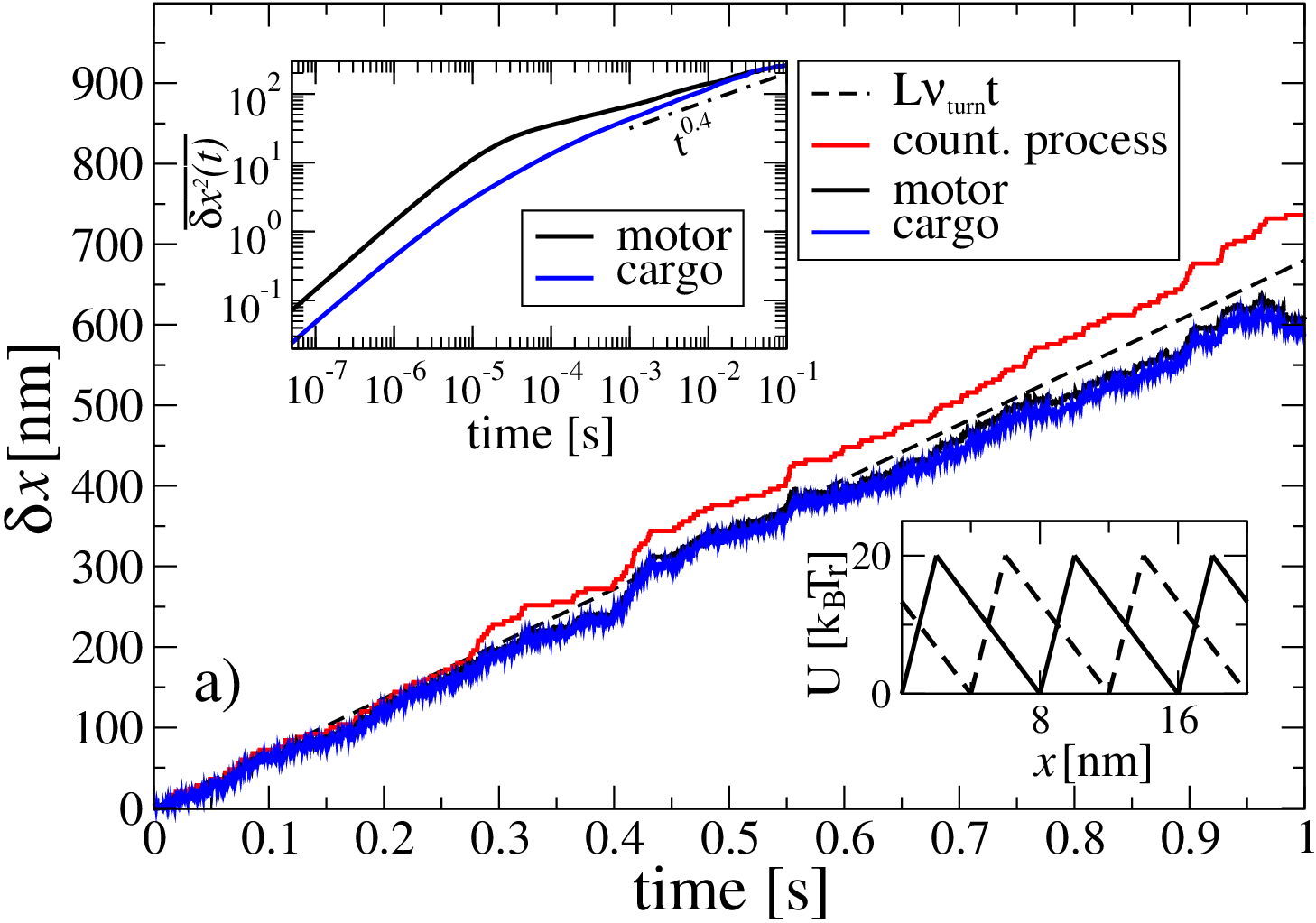} \hfill
      \includegraphics*[width=3.5 in]{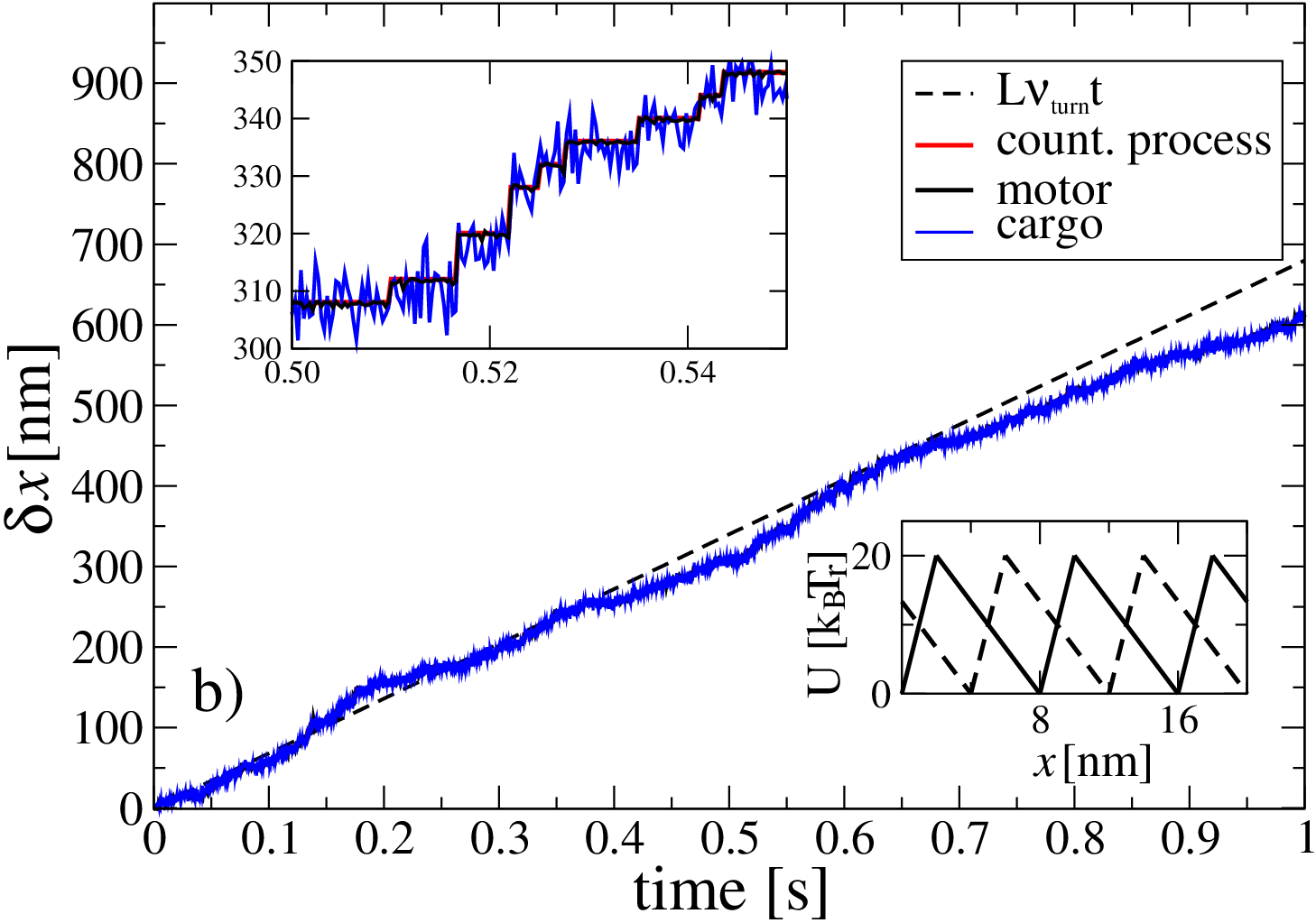}
      \caption{}
      \label{Fig1}
   \end{center}
\end{figure}

\begin{figure}
   \begin{center}
      \includegraphics*[width=3.5 in]{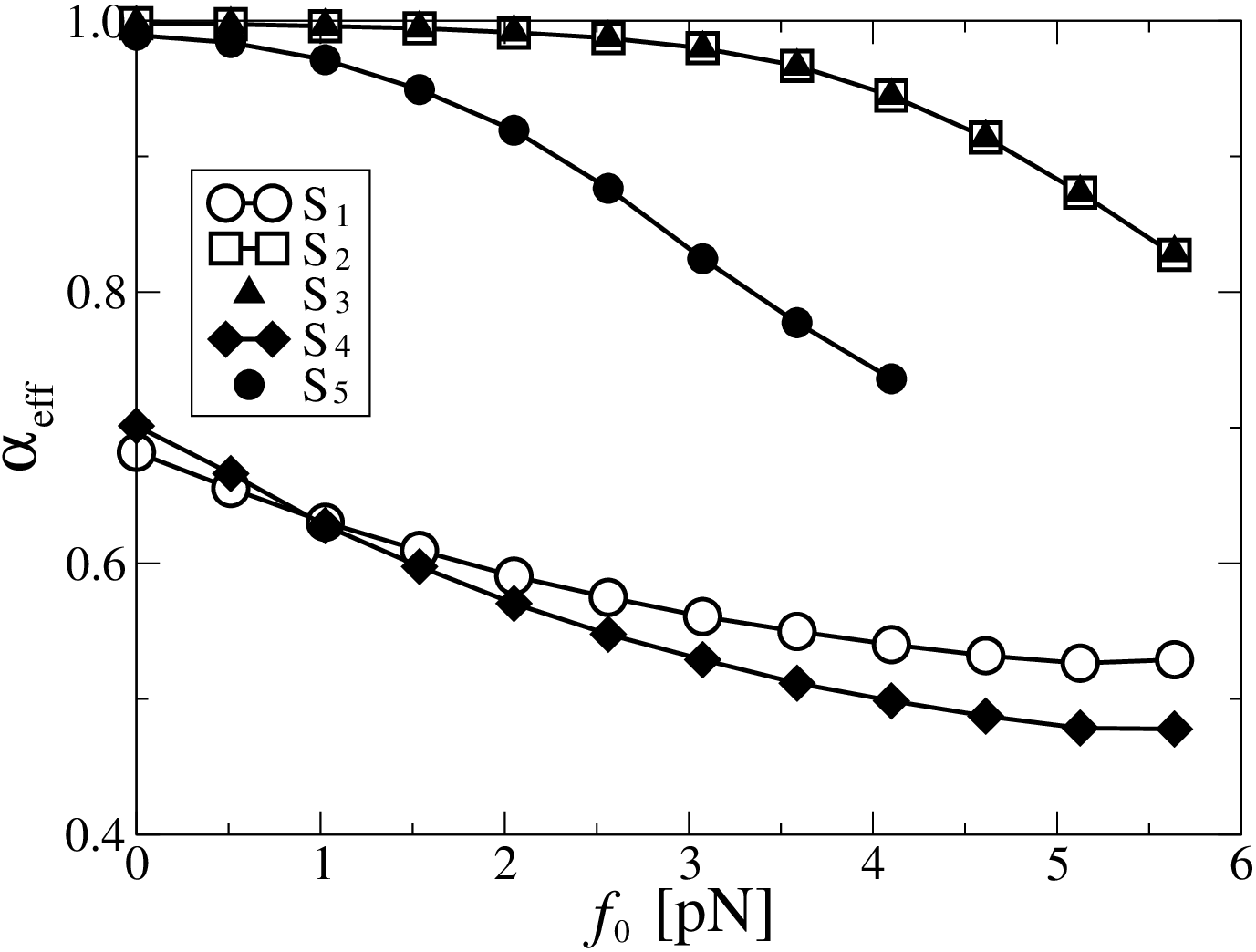} 
      \caption{}
      \label{Fig2}
   \end{center}
\end{figure}

First, we illustrate a single trajectory realization of anomalous transport for the set $S_1$ in Fig.
\ref{Fig1}, a. The upper inset in this figure shows the coupled diffusion of the cargo and the motor off
the microtubule track. Without any coupling, the motor particle diffuses normally on any time scale
within the considered model, and the cargo initially diffuses normally as well, but  then it subdiffuses
until the time scale $\tau_{\rm max}$ is reached. Here, a time-averaging  of the particle position
variances $\delta x^2(t|t')=[x(t+t')-x(t')]^2$ over the  corresponding single trajectories
(time-averaging over sliding $t'$ within a time window ${\cal T}$) is done, $\overline{\delta
x^2(t)}=\frac{1}{{\cal T}-t}\int_0^{{\cal T}-t}\delta x^2(t|t')dt'$ ($t\ll {\cal T}$).  It coincides with
the ensemble-average$\langle \delta x^2(t)\rangle $ since the considered viscoelastic  diffusion is
ergodic \cite{Goychuk09}. As it is clearly seen from  Fig. \ref{Fig1}, a, the coupled motor and cargo
subdiffuse together  $\langle \delta x^2(t)\rangle \propto t^{\alpha}$ after some transient time.  In
other words, subdiffusing particle enslaves the normally diffusing one, when the last one  is passive and
not empowered by trapping and  pulsing potential. There is no mean displacement of this complex on
average. However, when the motor is attached to the track, it processively steps in the direction defined
by the potential asymmetry, and the (freely) subdiffusing cargo cannot withstand. It follows to  the
winning motor. Initially, the motor steps are perfectly tight to the potential fluctuations (notice that
a particular realization of the counting process in units of $L/2$ in this figure is faster than the
average value corresponding to the optimal transport distance $d=L\nu_{\rm turn}^{(1)}t$, cf. the broken
line).  After some initial time one can clearly see backsteps of the motor and cargo is always
fluctuating around the motor position. However, it lags somewhat behind the motor on average. 
It is not obvious from this figure that the transport is anomalously slow and not just corresponds to
some suboptimal mean motor velocity $v<v_{\rm opt}=L\nu_{\rm turn}$. However, the transport is anomalous
indeed,  $\langle \delta x(t)\rangle \propto t^{\alpha_{\rm eff}}$, as it can be deduced after averaging
over 1000 different trajectory realizations. This allows to deduce the anomalous transport exponent
$\alpha_{\rm eff}$. As already verified in our previous studies of viscoelastic subdiffusive dynamics in
periodic nonlinear potentials, such  $\alpha_{\rm eff}$ is generally time-dependent.  However, it 
relaxes to a long-time limiting value in the course of time, which is displayed in Fig. \ref{Fig2} as
function of load $f_0$ for different sets of parameters.  For $S_1$ and $f_0=0$ in Fig. \ref{Fig1}, a, 
$\alpha_{\rm eff}\approx 0.7$. This explains the origin of superdiffusive exponent $\beta=1.4$, in spite
of the low value of free subdiffusion exponent  $\alpha=0.4$ in \citep{Robert}.  However, for a smaller
particle (or rather for a smaller $\eta_{\alpha}$ in our model), the case $S_2$, the transport becomes
normal at the same flashing frequency and without biasing back  load, $f_0=0$, see in Fig. \ref{Fig1}, b.
Interestingly,  the corresponding single trajectory  realization in Fig. \ref{Fig1}, b does not
correspond to a larger transport distance at $t=1$ s, maximal in this figure, as compare with anomalous
transport in Fig. \ref{Fig1}, a. This is simply because the number of turnovers done until this time is
larger in Fig. \ref{Fig1}, a than in Fig. \ref{Fig1}, b, for the particular realizations presented. This
reflects statistical variation of the corresponding counting process, or, in physical terms, stochastic
nature of  single motor proteins. A perfect synchronization between the potential switches and the motor
steps makes the orange line corresponding to the potential switches barely visible in 
Fig. \ref{Fig1}, b. One
can see also in the upper inset that cargo fluctuates around the motor symmetrically. It does not lag
behind the motor on average, like in Fig. \ref{Fig1}, a.  Furthermore, a much softer linker does not
change  qualitatively the results, as Fig. \ref{Fig2} demonstrates, though anomalous transport is
stronger affected. Unexpectedly, $\alpha_{\rm eff}$ is slightly larger for a softer linker at $f_0=0$ in
the anomalous regime. This is because we derived $\alpha_{\rm eff}$ from the position  of the motor,
rather than cargo and the enlarged mean distance between the motors and their cargos was still not
completely relaxed to a stationary value for $S_4$ in Fig. \ref{Fig2}. Moreover, the anomalous transport 
can become normal if the turnover frequency is reduced, see the results for $S_5$ in Fig. \ref{Fig2}.
These results show that motors can realize both normal and anomalous transport in viscoelastic  cytosol
of living cells, where the large particles subdiffuse on the time scale from milliseconds to seconds.
Which regime will be realized depends in particular on the particle size  (or fractional friction
coefficient  $\eta_{\alpha}$) and the enzyme turnover frequency $\nu_{\rm turn}$, but only weakly depends
on the linker rigidity. This presents one of the important results of our work. Furthermore, if to apply
a counter-force  $f_0>0$, the anomalous transport regime becomes promoted, cf. in Fig. \ref{Fig2}. This
effect is primarily due to reduction of the binding potential amplitude  $U_0$. Anomalous transport will
also emerge immediately for the studied parameters if to decrease  $U_0$ essentially, e.g. to $10\;k_B
T_r$.

\begin{figure}
   \begin{center}
      \includegraphics*[width=3.5 in]{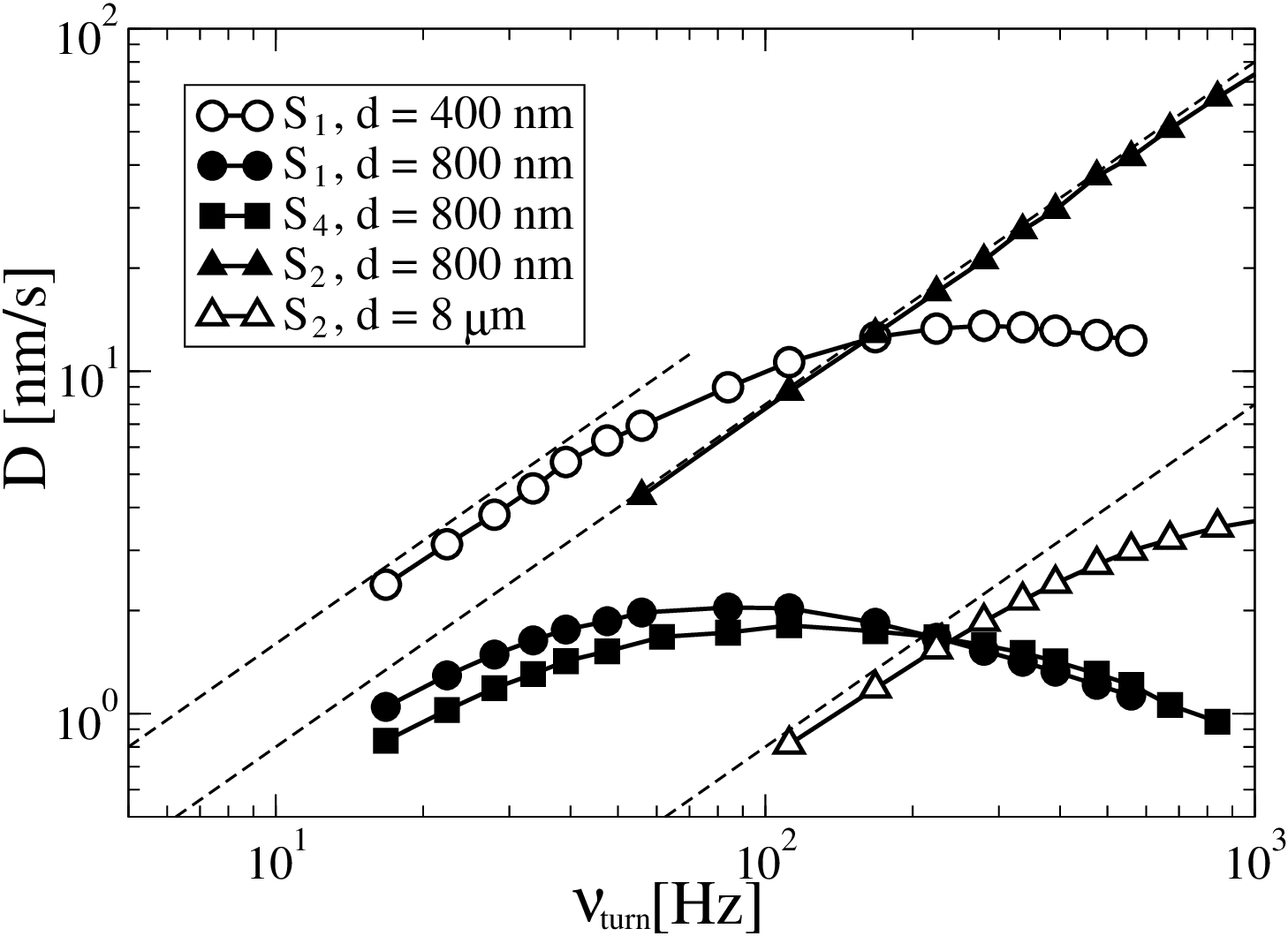} 
      \caption{}
      \label{Fig3}
   \end{center}
\end{figure}

The transport efficiency of  molecular motors in viscoelastic cytosol can be almost
perfect, despite subdiffusion, as Fig. \ref{Fig3} demonstrates for $S_2$ and realistic turnover frequencies 
(lower than 200 Hz) even for large distances like 8 $\mu m$. The discussed power stroke mechanism
can perfectly overcome subdiffusive resistance of medium, if the cargo size is not too large. 
In anomalous transport regime
with strongly decreased transport efficiency 
there exists but an optimal turnover frequency, which depends on the delivery distance.
Interestingly, this optimization does not depend manifestly on the rigidity of 
linker, see in Fig. \ref{Fig3}. However, anomalous transport on a stronger linker
is yet more efficient for realistic turnover frequencies. And this agrees with our intuition.

\begin{figure}
   \begin{center}
      \includegraphics*[width=3.5 in]{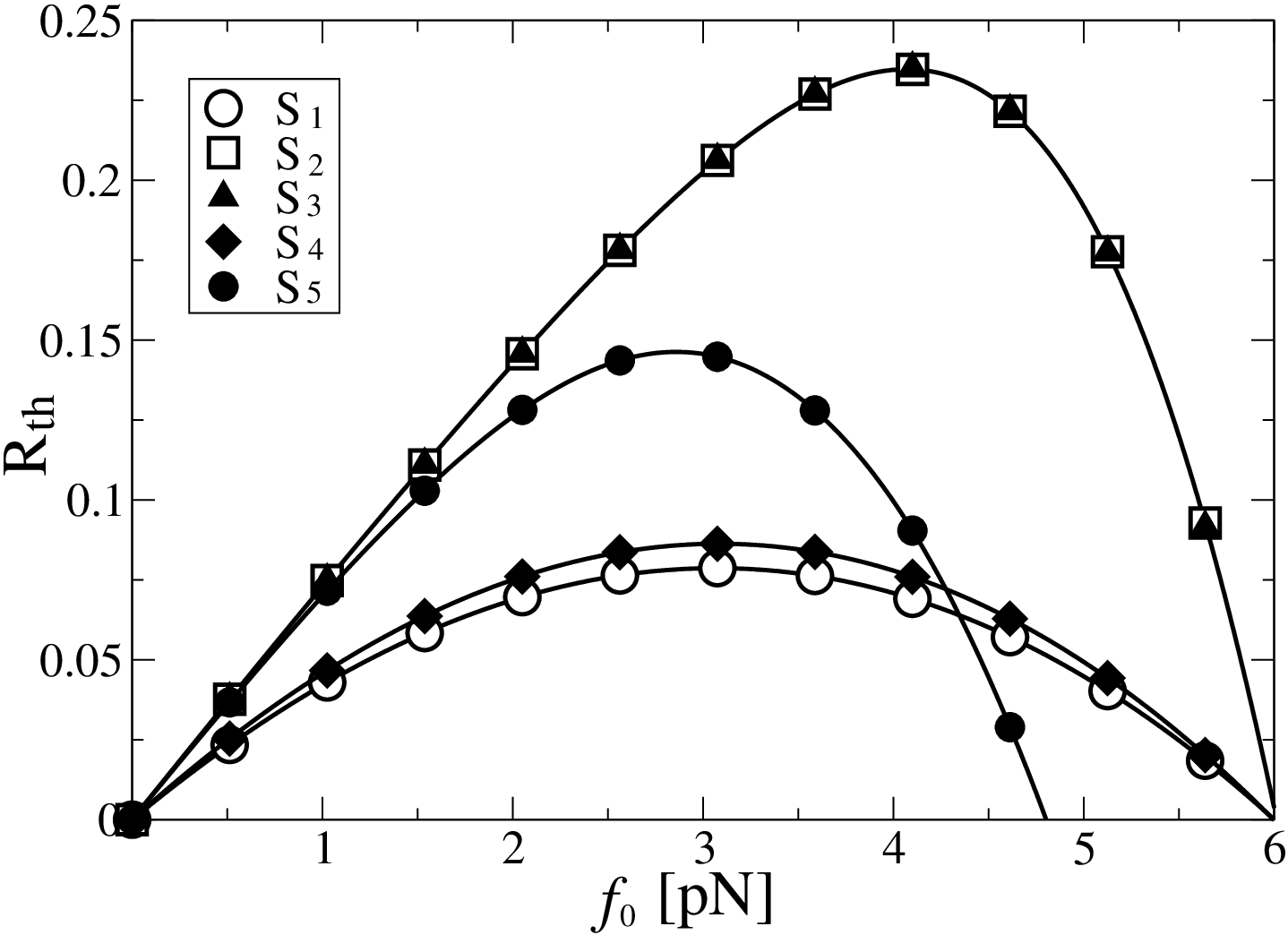} 
      \caption{}
      \label{Fig4}
   \end{center}
\end{figure}

Thermodynamic efficiency becomes strongly affected by the fact that
an increase of $f_0$ promotes anomalous transport regime, see in Fig. \ref{Fig2}.
Even if the transport was normal at $f_0$, it becomes anomalous under sufficiently 
strong external load $f_0$.
Hence thermodynamic efficiency starts to depend on time and 
it vanishes asymptotically, whenever the transport
is anomalous. However, it can be substantially large
even at large times such as $t_{\rm max}\sim 3$ s in Fig. \ref{Fig4}, 
where it is still about 23\% at maximum, see results for
$S_2,S_3$, where thermodynamic efficiency practically does not depend on the linker rigidity.
For low efficient anomalous transport, the dependence of $R_{\rm th}$ on load is approximately
parabolic with maximum at $f_0^{opt}=f_0^{stop}/2$. Counterintuitively, it is slightly larger
for softer linker in Fig. \ref{Fig4}. This is because the useful work has been defined as one done
by motors against $f_0$, and not  by cargos, and the motors step ahead their cargos
 on slightly larger distances for a softer linker. The maximal  loading, or stalling
force $f_0^{stop}$ is time-independent. It also does not depend on the cargo size.
However, it strongly depends on the flashing frequency (see in Fig. \ref{Fig4}), and also
on the potential amplitude and temperature, see in Fig. \ref{Fig5} for a fixed $\nu_{\rm turn}=85$ Hz.
Numerical results shows that for $U_0>U_m(\nu_{\rm turn})T/T_r$, 
\begin{eqnarray}\label{stop_force}
f_0^{stop}(T,U_0,\nu_{\rm turn})
\approx \frac{4}{3L}\left(U_0-U_m(\nu_{\rm turn})\frac{ T}{T_r} \right),
\end{eqnarray}
where fit to numerical data yields $U_m\approx 11.2\;k_BT_r$ at $\nu_{\rm turn}=85$ Hz. 
The corresponding stalling force $f_0^{stop}\approx 6$ pN for $U_0=20\;k_BT_r$, but
it is smaller for $\nu_{\rm turn}=17$ Hz, see in Fig. \ref{Fig4}.
From this one can conclude that a reasonably strong motor requires binding potential amplitudes
larger than ten $k_BT$. The result in Eq. \ref{stop_force} allows for a physical
interpretation upon introduction of an effective free energy barrier $F_0(T)=U_0-TS_0$
with an effective ``entropy'' $S=U_m(\nu_{\rm turn})/T_r$. Then, 
$f_0^{stop}(T,U_0,\nu_{\rm turn})\approx 4F_0(T)/3L$, for positive $F_0$.

\section*{Discussion and Conclusions}

In this work, we elaborated on a model of active molecular transport realized by molecular 
motors in the case when their cargos are subdiffusing when left alone. 
Subdiffusion is described by a Generalized Langevin Equation with a memory kernel
which scales in accordance with a power law between two memory cutoffs. Subdiffusion 
is realized until the long-time memory cutoff $\tau_{\rm max}$ is reached, and the time range 
of subdiffusion is determined by this time and an effectively enhanced relative (with respect 
to water) cytosol viscosity $\zeta_{\rm rel}$, which depends on the particle size.  The effective
viscosity defines asymptotically 
normal diffusion regime, and initially diffusion is also normal.
Subdiffusion occurs on the time scale between  $\tau_{\rm in}=\tau_{\rm max}/\zeta_{\rm rel}^{1/(1-\alpha)}$
and $\tau_{\rm max}$, and it can occur over about 6 to 7 time decades for 
$\zeta_{\rm rel}\sim 10^4$, or about 5 time decades for
$\zeta_{\rm rel}\sim 10^3$, and $\alpha=0.4$.  
Such transient subdiffusion allows for a nice multidimensional 
Markovian embedding with a well-controlled
accuracy of approximation by using a set of 
overdamped Brownian quasi-particles elastically attached 
to cargo on harmonic springs, with spring constants and frictional coefficients 
obeying a fractal scaling. 
The molecular motor is described by a variant of standard model of flashing Brownian motors with 
a periodic saw-teeth potential randomly fluctuating between two realizations differing by phase, 
so that two potential
fluctuations corresponding to one completed enzyme cycle can promote the motor by one spatial period. 
This model describes, in particular,  a perfect power stroke ratchet transport in the case
of highly processive molecular motors with binding potential amplitude exceeding a minimal height
of the order of $10\;k_BT_r$, see in Fig. \ref{Fig5}, with $U_0=20\;k_BT_r$ in this work. 
A perfect transport regime with maximal
transport efficiency emerges when the motion of motor is locked to
the potential fluctuations caused by the change of the internal motor state.
The cargo is coupled elastically to the motor. This model differs profoundly from two
previous modeling routes to describe active anomalous transport and diffusion in living cells.
In a simplest model, the influence of motors is modeled by random force exhibiting weakly decaying
correlations. It cannot, however, explain the origin of the observed superdiffusion in the
cells with the passive subdiffusion exponent less or equal $\alpha=0.5$, like 
$\alpha=0.4$ detected in Refs. \citep{Robert,Bruno} and used also in this work.
A better model has been introduced recently in \citep{GKhMetz13a}. It assumes, however, a perfectly rigid 
linker between the motor and the cargo, like most models of molecular motors do, 
so that the cargo and the motor make one effectively subdiffusing (when it is left alone)
Brownian particle moving in a flashing potential. That one is a model of anomalous Brownian motors.
The crucial point which it explains is how one and the same motor, in the same cell
can realize both normal and anomalous transport. The occurrence of particular transport
regime depends on the binding potential
amplitude, fractional frictional strength $\eta_{\alpha}$ (depending on the cargo size),
loading external force $f_0$,
and enzyme turnover frequency. The effective transport exponent $\alpha_{\rm eff}$
can vary from $\alpha$ to one, and this can easily explain the emergence
of anomalously fast diffusion with $\beta=2\alpha_{\rm eff}$ mediated by motors in the cells 
with $\alpha\leq 0.5$. Strikingly enough, the transport can be not only normal (in agreement with most
experiments), but also reflect an almost perfect synchronization between the enzymatic 
turnovers and the motor's stepping along microtubule. This can be rationalized within a power-stroke mechanism 
and explains how  a power-stroke like operation can beat subdiffusion.

The major advance of this work is the clarification of how this picture is modified upon considering
realistically soft linkers between the motor (operating normally in the absence of cargo,
in neglection of viscoelastic memory effects) and the subdiffusing cargo. It turns out that all the 
major features revealed in \citep{GKhMetz13a} survive, qualitatively all the results look similar.
However, anomalous
transport regime can emerge already for turnover frequencies less than 100 Hz (about maximal
frequency for kinesin motors) and the cargo sizes about $a=300$ nm. We believe that
these new results will inspire an experimental verification, because they should
survive also in more complicated models of molecular motors operating in viscoelastic
cytosol. We hope that this can be done in a nearest future.



\clearpage
\section*{Figure Legends}
\subsubsection*{Figure~\ref{Fig1}.}
Positions of motor (black line) and cargo (blue line) versus time for a single trajectory 
realization in the case of anomalous transport 
(part a, set $S_1$, $\tilde \eta_{\rm eff}=3\cdot 10^4$, 
see text for the parameters corresponding to various sets $S_i$) and
normal transport (part b, set $S_2$, 
$\tilde \eta_{\rm eff}=3\cdot 10^3$). Turnover frequency $\nu_{\rm turn}=85$ Hz.
The realizations of counting process (number of potential switches multiplied with potential 
half-period $L/2$) are depicted by orange lines (difficult to detect in part \textbf{b} because of a perfect
synchronization). The broken black lines depict the dependence of the averaged 
(over many trajectory realizations) position of motor on time
in the case of a perfect synchronization (ideal power stroke like mechanism).
Upper inset in the part \textbf{a}, shows diffusion of coupled motor and cargo in the absence
of binding potential. The position variances have been obtained using a corresponding single-trajectory
time averaging, as described in the text. The upper inset in the part \textbf{b} magnifies a part of
motor and cargo trajectories making the step-wise motion of motor obvious. It is perfectly synchronized
with the potential switches. Cargo randomly fluctuates around the motor position. The lower insets
show two conformations of binding potential.

\subsubsection*{Figure~\ref{Fig2}.}

Effective transport exponent $\alpha_{\rm eff}$ as function of loading force $f_0$ for various
sets of parameters.

\subsubsection*{Figure~\ref{Fig3}.}

Transport delivery efficiency $D$ as function of turnover frequency 
$\nu_{\rm turn}$ for different sets of parameters
and different delivery distances $d$. Broken lines correspond
to ideal power-stroke dependence $D_{\rm ideal}=L^2\nu_{\rm turn}/d$, to compare
with.

\subsubsection*{Figure~\ref{Fig4}.}

Thermodynamic efficiency as function of loading force $f_0$ at the end of simulations 
(corresponding to $t_{\rm max}=2.94$ s)
for different sets.

\begin{figure}
   \begin{center}
      \includegraphics*[width=3.5 in]{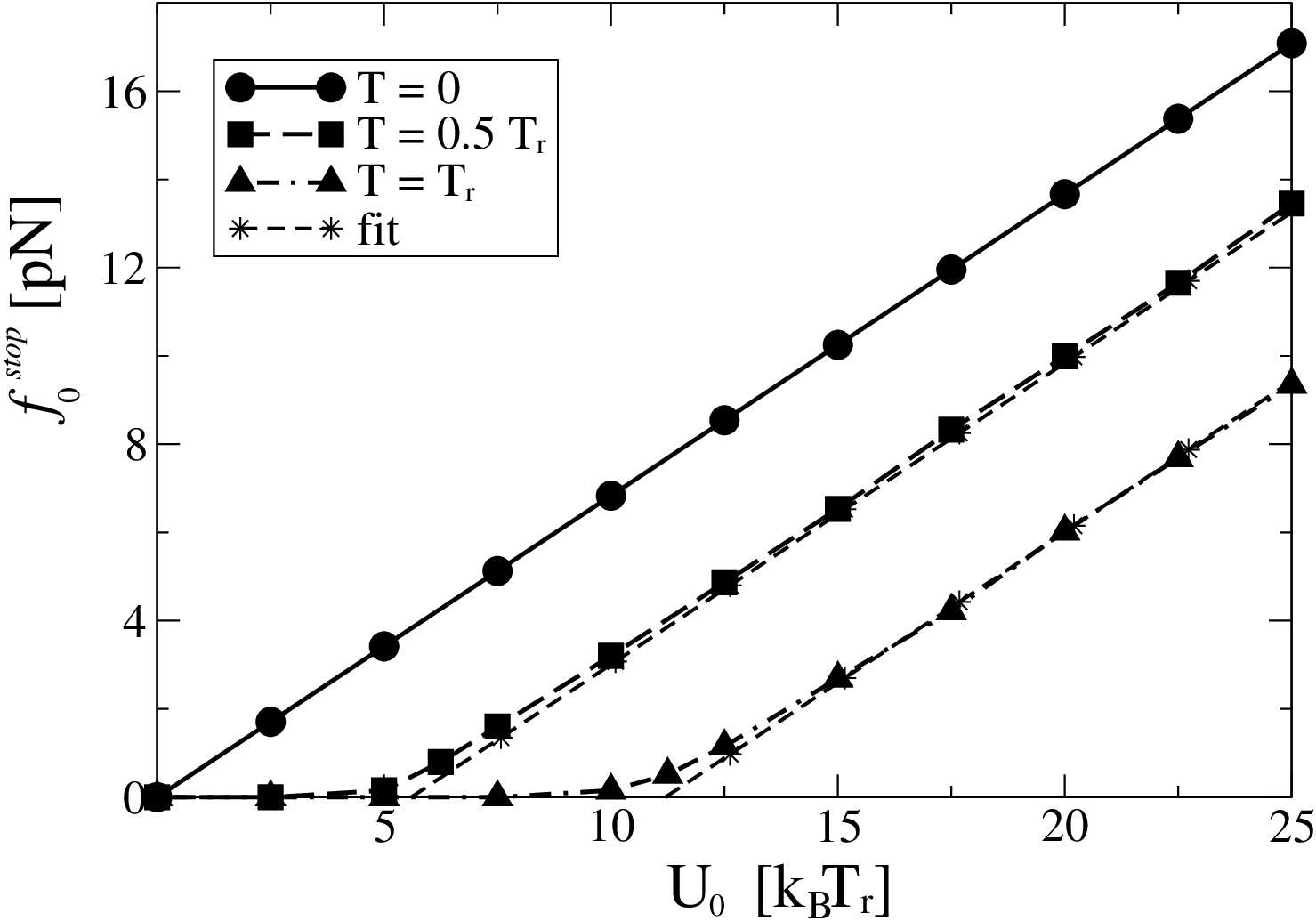} 
      \caption{}
      \label{Fig5}
   \end{center}
\end{figure}

\subsubsection*{Figure~\ref{Fig5}.}

 Stalling force as function of barrier height $U_0$ at different temperatures 
 and fixed $\nu_{\rm turn}=85$ Hz. Numerical results are compared with 
 an analytical fit by Eq. \ref{stop_force} for $U_0> U_m T/T_r$.

\clearpage

\textbf{Acknowledgments}\\
Support of this research by the German Research Foundation, Grants
GO 2052/1-1 and GO 2052/1-2, as well as funding from the Academy of Finland
(FiDiPro scheme) are gratefully acknowledged. \\

\end{document}